\documentclass[11pt,a4page]{article}
\usepackage{amsfonts}
\usepackage{amsmath}
\usepackage{amsthm}
\usepackage{amssymb}
\usepackage{amscd}

\def\be{\begin{equation}}
\def\ee{\end{equation}}

\def\bea{\begin{eqnarray}}
\def\eea{\end{eqnarray}}

\def\a{\alpha}
\def\b{\beta}
\def\c{\gamma}
\def\C{\Gamma}
\def\e{\epsilon}
\def\si{\sigma}
\def\Si{\Sigma}

\begin{document}

\begin{flushright}
DAMTP-2002-13
\end{flushright}

\vspace{4cm}

\begin{center}
{\LARGE {\bf Dirac equation on a $G_2$ manifold}}
\vspace{1cm}

Sean A. Hartnoll \\
\vspace{0.3cm}

{\it DAMTP, Centre for Mathematical Sciences, Cambridge University,\\
Wilberforce Road, Cambridge CB3 0WA, UK.}
\vspace{0.5cm}

\noindent e-mail: S.A.Hartnoll@damtp.cam.ac.uk
\end{center}
\vspace{1cm}

\begin{abstract}
We find a large family of solutions
to the Dirac equation on a manifold
of $G_2$ holonomy asymptotic to a cone over $S^3 \times S^3$,
including all radial solutions. The behaviour
of these solutions is studied as the manifold
developes a conical singularity. None of the solutions found are both localised
and square integrable at the origin. This result is consistent with the
absence of chiral fermions in M-theory on the conifold over $S^3\times S^3$.
The approach here is complementary to previous analyses using dualities
and anomaly cancellation.
\end{abstract}

\pagebreak
\pagenumbering{arabic}

\section{Introduction}

Eleven dimensional supergravity cannot give a chiral four dimensional theory upon compactification on a smooth
manifold \cite{wit1}. One way to resolve this discrepancy with experiment is by compactifying on
a manifold with boundary \cite{hw1,hw2}. Recently, work has centred on obtaining chiral fermions by the alternative
approach of compactifying on singular manifolds. As supergravity is no longer valid near the singular points, the
arguments make essential use of M-theory. Given that M-theory remains unformulated, this means they work via dualities
with string theories. Approaches include duality with intersecting D6-branes and orientifolds of type IIA string theory
\cite{aw,csu1,csu2} and fibrewise duality with Heterotic string theory compactified on certain Calabi-Yau manifolds \cite{acw}.
The duality results are corroborated by Witten's analysis of anomaly cancellation at conical singularities \cite{wit2}.

A recurrent theme in these arguments is compactification on manifolds of $G_2$ holonomy (\cite{jo} for
mathematical background, \cite{gub} for a pedagogical introduction). This is unsurprising as
supersymmetry plays an important r\^{o}le in duality arguments and $G_2$ holonomy is the condition on the internal
seven dimensional manifold for compactification from eleven to four dimensions to preserve $\mathcal{N}=1$
supersymmetry. The preservation of $\mathcal{N}=1$ supersymmetry is phenomenologically appealing and
$G_2$ manifolds were considered in early work on Kaluza-Klein supergravity \cite{dnp,duff}. The possibility
of further obtaining chiral theories with non-abelian gauge symmetry via M-theory is even more phenomenologically
interesting.

Parallel to the renewed interest in the physics of $G_2$ holonomy compactifications,
there has been a lot of recent work on constructing and classifying $G_2$ metrics.
All the examples known are cohomogeneity-one metrics on noncompact spaces
that are asymptotic to generalised cones \cite{cglp1, cgllp, cglp3, cglp4, cglp5, cglp6, bggg, bh, bds, cs}.
The metrics constructed are generalisations of the metics found a decade ago by \cite{bs,gpp}.
They are asymptotic to cones over $SU(3)/U(1)\times U(1)$, $\mathbb{CP}^3$ and $S^3 \times S^3$.
No explicit metrics are known on compact $G_2$ manifolds because these cannot have
continuous isometries except for trivial $S^1$ factors. The appearance of chiral fermions is a phenomenon
localised near the conical singularity and so considering noncompact manifolds should not be problematic.

The original argument against obtaining chiral theories from eleven dimensions \cite{wit1} is circumvented by the fact that
the non-abelian gauge groups under which the chiral fermions are charged do not come from isometries of the
internal manifold, but from massless branes wrapping collapsed homology cycles in the internal manifold \cite{wit3,bvs}.
It would be nice to see these new massless fermionic degrees of freedom emerge as homology cycles collapse directly
within an eleven dimensional framework, without using dualities. Further, for the $G_2$ manifold asymptotic to a cone
over $S^3\times S^3$, call it $X$, the situation regarding chiral fermions is not yet understood, although other aspects
of M-theory on this space have been studied in some detail (e.g. \cite{aw, amv}). There are various reasons for this:
{\it (a)} The anomaly analysis of \cite{wit2} is not applicable as for this space $H^2(X;U(1))=0$. {\it (b)}
The $G_2$ manifolds considered in \cite{acw} were $\mathbb{R}^3$ bundles over a self-dual four-fold whilst
$X$ is an $\mathbb{R}^4$ bundle over $S^3$. {\it (c)} Whilst the cones over $SU(3)/U(1)\times U(1)$
and $\mathbb{CP}^3$ were interpreted in \cite{aw} as the M-theory lift of intersecting D6-branes, the
analogous interpretation for the $S^3 \times S^3$ case is not clear.

The line of attack in this work will be to look for solutions to the massless Dirac equation on manifolds on $G_2$ holonomy and
see how the behaviour of the solutions changes as the manifold developes a conical singularity. We use the Dirac equation
and not the Rarita-Schwinger spin-3/2 equation because the degrees of freedom we are interested in do not
originate from supergravity but from wrapped branes. It turns out that the $G_2$ manifold on which
the Dirac equation is most tractable, allowing us to find all radial solutions and several non-radial solutions,
is precisely the one asymptotic to a cone over $S^3\times S^3$. Thus the information we gain through this
approach is nicely complementary to the information already obtained
through dualities. Solving the Dirac equation on other $G_2$ manifolds
could also be interesting, but is not done here.

In \S2 we write down the massless Dirac equation on the $G_2$ manifold and find all radial solutions by directly
solving the equation. In \S3 we find some non-radial solutions by relating solutions of the massless Dirac equation
to solutions of the massless Klein-Gordon equation via the covariantly constant spinor on the $G_2$ manifold.
The qualitative behaviour of the solutions does not change in the singular limit and none of the solutions
have the property of being both localised at the origin and square integrable at the origin, properties which
we should expect for degrees of freedom corresponding to the chiral fermions.
In \S4 we discuss these results in the context of chiral fermions appearing in singular M-theory compactifications.
We suggest that our results are consistent with the absence of chiral fermions arising in the singular
limit of M-theory compactified on a $G_2$ holonomy manifold asymptotic to a cone over $S^3 \times S^3$.

\section{Dirac equation and all radial solutions}

We will work with the simplest $G_2$ holonomy metric asymptotic to a cone over $S^3\times S^3$ which
has metric \cite{bs,gpp}
\be\label{eq:metric}
ds^2 = \a(r)^2 dr^2 + \b(r)^2 (\si^i - \frac{1}{2}\Si^i)^2 + \c(r)^2 \Si^i \Si^i ,
\end{equation}
where the $\si^i$ and the $\Si^i$ are left invariant one-forms on the two copies of $S^3$, that
is $d\si^i = -\frac{1}{2} \e^{i j k} \si^j \wedge \si^k$ and $d\Si^i = -\frac{1}{2} \e^{i j k} \Si^j \wedge \Si^k$.
The radial functions are
\be\label{eq:fns}
\a(r)^2 = (1-\frac{r_0^3}{r^3})^{-1} ; \quad \b(r)^2 = \frac{1}{9} r^2 (1-\frac{r_0^3}{r^3}) ; \quad \c(r)^2 = \frac{1}{12} r^2 .
\end{equation}
The range of $r$ is $r_0 \leq r < \infty$. It is useful to consider the vielbeins
\be\label{eq:viel}
\hat{e}^0 = \a dr ; \quad \hat{e}^{\hat{i}} = \b (\si^i - \frac{1}{2} \Si^i) ; \quad \hat{e}^i = \c \Si^i .
\end{equation}
We will use tangent space indices $i,j,k=1,2,3$ and $\hat{i},\hat{j},\hat{k}=\hat{1},\hat{2},\hat{3}=4,5,6$ and
also $a,b,c... = 0...6$. Spacetime indices will be $\mu, \nu ...$.

The massless Dirac equation corresponding to the metric (\ref{eq:metric}) can be read off from the calculation of the
connection one-form in \cite{gpp},
\bea\label{eq:dirac}
\C^a D_a \psi & = & \C^a \left(e_a - \frac{1}{4}\omega^{bc}_{\quad a} \C_{bc} \right) \psi \nonumber \\
& = & \left(\C^a e_a + \frac{\b^{\prime}}{2\a\b} \C^{\hat{i}} \C_{0 \hat{i}}
+  \frac{\c^{\prime}}{2\a\c} \C^i \C_{0 i} +
\e^{i j k} \C_{\hat{k}} (\frac{1}{8\b} \C_{\hat{i}\hat{j}} + \frac{\b}{32\c^2}\C_{i j}) \right. \nonumber \\
&  & \left. +  \e^{i j k} \C_k (\frac{1}{8\c} \C_{\hat{i}\hat{j}} - \frac{\b}{16\c^2}\C_{i \hat{j}}
+ \frac{1}{8\c}\C_{i j} ) \right) \psi = 0 ,
\eea
where as usual $\C_{ab} = \frac{1}{2} \left[\C_a ,\C_b \right]$ and $e_a$ are vector fields dual to the
vielbeins (\ref{eq:viel}), $<e^b, e_a> = \delta^b_a$. We use a convention for the gamma matrics such that
the Clifford algebra is $\{ \C_a ,\C_b \} = -2\delta_{ab}$.

To solve this equation we need to choose a representation for the Euclidean
Clifford algebra in seven dimensions. For this algebra there is a
representation in which all the matrics are real. It is given by \cite{cdfn,gg}
\be
(\C_a)_{s t} = c_{a s t} + \delta_{a s} \delta_{t 7} - \delta_{a t} \delta_{s 7} ,
\end{equation}
where the spinor indices $s,t = 0...7$ and $c_{a s t}$ are zero if $s=7$
or $t=7$ and otherwise are the totally antisymmetric octonion
structure constants, with
\be
c_{016} = c_{052} = c_{043} = c_{142} = c_{135} = c_{236} = c_{456} = 1 .
\end{equation}
In this representation, the Majorana or reality condition is particularly simple
\be\label{eq:real}
\psi^{*} = \psi .
\end{equation}
The Dirac operator (\ref{eq:dirac}) takes Majorana (real) spinors to Majorana
spinors. Therefore by expressing an arbitrary spinor in terms of two
Majorana spinors
\be
\psi = \psi^{(1)} + i \psi^{(2)} ,
\end{equation}
we may consider the Dirac equation for $\psi^{(1)}$ and $\psi^{(2)}$ separately.

For the remainder of this section we take $\psi$ to be Majorana. 
We will thus obtain half of the solutions. However, the
other half will be of exactly the same form but with a pure imaginary
rather than real coefficient.
Writing out $\psi$ in components
\be\label{eq:realspin}
\psi = \left( f_1 , f_2, f_3, f_4, f_5, f_6, f_7, f_8 \right) ,
\end{equation}
where $f_1 ... f_8$ are real functions.
The Dirac equation
thus becomes eight coupled equations for eight real functions. These equations
are only tractable if one makes the radial ansatz
\be
e_i \psi = e_{\hat{i}} \psi = 0 ,
\end{equation}
for $i=1,2,3$. That is, we restrict the spinor to depend on $r$ only.

Substituting (\ref{eq:realspin}) into the Dirac equation (\ref{eq:dirac}) one obtains the eight equations
\bea
r \frac{\partial f_8}{\partial r} + \frac{f_3}{2} -
\sqrt{3}\sqrt{\frac{r^3}{r^3-r_0^3}} f_6 - \frac{1}{4} \left(
\frac{10r^3-r_0^3}{r^3-r_0^3} \right) f_1 + \frac{1}{4} \left(
\frac{12r^3-3r_0^3}{r^3-r_0^3} + 4 \sqrt{3} \sqrt{\frac{r^3}{r^3-r_0^3}}
\right) f_8 & = & 0 , \nonumber \\
r \frac{\partial f_3}{\partial r} + \frac{f_8}{2} -
\sqrt{3}\sqrt{\frac{r^3}{r^3-r_0^3}} f_1 + \frac{1}{4} \left(
\frac{10r^3-r_0^3}{r^3-r_0^3} \right) f_6 + \frac{1}{4} \left(
\frac{12r^3-3r_0^3}{r^3-r_0^3} - 4 \sqrt{3} \sqrt{\frac{r^3}{r^3-r_0^3}}
\right) f_3 & = & 0 , \nonumber \\
r \frac{\partial f_6}{\partial r} - \frac{f_1}{2} -
\sqrt{3}\sqrt{\frac{r^3}{r^3-r_0^3}} f_8 + \frac{1}{4} \left(
\frac{10r^3-r_0^3}{r^3-r_0^3} \right) f_3 + \frac{1}{4} \left(
\frac{12r^3-3r_0^3}{r^3-r_0^3} + 4 \sqrt{3} \sqrt{\frac{r^3}{r^3-r_0^3}}
\right) f_6 & = & 0 , \nonumber \\
r \frac{\partial f_1}{\partial r} - \frac{f_6}{2} -
\sqrt{3}\sqrt{\frac{r^3}{r^3-r_0^3}} f_3 - \frac{1}{4} \left(
\frac{10r^3-r_0^3}{r^3-r_0^3} \right) f_8 + \frac{1}{4} \left(
\frac{12r^3-3r_0^3}{r^3-r_0^3} - 4 \sqrt{3} \sqrt{\frac{r^3}{r^3-r_0^3}}
\right) f_1 & = & 0 , \nonumber \\
r \frac{\partial f_7}{\partial r} - \frac{1}{4} \left(
\frac{8r^3+r_0^3}{r^3-r_0^3} \right) f_2 + \frac{1}{4} \left(
\frac{12r^3-3r_0^3}{r^3-r_0^3} + 8 \sqrt{3} \sqrt{\frac{r^3}{r^3-r_0^3}}
\right) f_7 & = & 0 , \nonumber \\
r \frac{\partial f_2}{\partial r} - \frac{1}{4} \left(
\frac{8r^3+r_0^3}{r^3-r_0^3} \right) f_7 + \frac{1}{4} \left(
\frac{12r^3-3r_0^3}{r^3-r_0^3} - 8 \sqrt{3} \sqrt{\frac{r^3}{r^3-r_0^3}}
\right) f_2 & = & 0 , \nonumber \\
r \frac{\partial f_5}{\partial r} + \frac{1}{4} \left(
\frac{8r^3+r_0^3}{r^3-r_0^3} \right) f_4 + \frac{1}{4} \left(
\frac{12r^3-3r_0^3}{r^3-r_0^3} + 8 \sqrt{3} \sqrt{\frac{r^3}{r^3-r_0^3}}
\right) f_5 & = & 0 , \nonumber \\
r \frac{\partial f_4}{\partial r} + \frac{1}{4} \left(
\frac{8r^3+r_0^3}{r^3-r_0^3} \right) f_5 + \frac{1}{4} \left(
\frac{12r^3-3r_0^3}{r^3-r_0^3} - 8 \sqrt{3} \sqrt{\frac{r^3}{r^3-r_0^3}}
\right) f_4 & = & 0 . \nonumber \\ 
\eea

Note that these equations form three
groups; for $\{f_2,f_7\}$, for $\{f_4,f_5\}$ and for $\{f_1,f_3,f_6,f_8 \}$.
To discuss the solutions, we need to distinguish the cases $r_0 \neq 0$
and $r_0 = 0$. In each case we obtain eight linearly independent solutions which, when combined with the
other eight solutions of the same form but pure imaginary coefficient, are all the radial solutions
to the Dirac equation (\ref{eq:dirac}).

\subsection{Solutions with $r_0 \neq 0$}

Define $\bar{r} = r - r_0$. We are interested in whether the solutions are square integrable, i.e. $L^2$, near the origin. For this
we need the determinant of the metric (\ref{eq:metric}) which near the origin is $\sqrt{g} \sim \bar{r}$ as
$\bar{r} \to 0$. The behaviour of the eight solutions is summarised in the following table.

\begin{table}[h]
  \begin{tabular}{|c|c|c|c|c|} \hline
Independent solns. & As $\bar{r} \to 0$ & As $\bar{r} \to \infty$ & $L^2$ at origin? & Localised? \\ \hline \hline
$\{f_2,f_7\}; \, \{f_4,f_5\}; \, \{f_1,f_3,f_6,f_8\} $ & $\sim \bar{r}^0$ & $\sim \bar{r}$ & Yes & No \\ \hline
$\{f_2,f_7\}; \, \{f_4,f_5\}; \, \{f_1,f_3,f_6,f_8\} $ & $\sim \bar{r}^{-3/2}$ & $\sim \bar{r}^{-7}$ & No & Yes \\ \hline
$\{f_1,f_3,f_6,f_8\}$ & $\sim \bar{r}^0$ & $\sim \bar{r}^0$ & Yes & No \\ \hline
$\{f_1,f_3,f_6,f_8\}$ & $\sim \bar{r}^{-3/2}$ & $\sim \bar{r}^{-6}$ & No & Yes \\ \hline
  \end{tabular}
\end{table}

Where the notation in the table means that, for instance, there is a
solution in which $f_4$ and $f_5$ go as a constant
near the origin and as $\bar{r}$ at infinity and all other radial functions are zero. The second and third columns show
only the asymptotic power-law behaviour, not the numerical coefficients, which in general are
not the same for the nonzero radial functions of each solution.
The general radial solution is a linear superposition of these eight solutions with complex coeffients.
None of the solutions are both $L^2$ at the origin and localised at
the origin. We are not interested in the behaviour at infinity,
although it is easy to see that none of the solutions are $L^2$ everywhere, in agreement with Lichnerowicz's theorem for
noncompact manifolds.

\subsection{Solutions with $r_0 = 0$}

The form of the determinant at the origin is now $\sqrt{g} \sim r^6$ as
$r \to 0$. The following table summarises the behaviour of the eight solutions.

\begin{table}[h]
  \begin{tabular}{|c|c|c|c|c|} \hline
Independent solns. & As $r \to 0$ & As $r \to \infty$ & $L^2$ at origin? & Localised? \\ \hline \hline
$\{f_2,f_7\}; \, \{f_4,f_5\}; \, \{f_1,f_3,f_6,f_8\} $ & $\sim r$ & $\sim r$ & Yes & No \\ \hline
$\{f_2,f_7\}; \, \{f_4,f_5\}; \, \{f_1,f_3,f_6,f_8\} $ & $\sim r^{-7}$ & $\sim r^{-7}$ & No & Yes \\ \hline
$\{f_1,f_3,f_6,f_8\}$ & $\sim r^0$ & $\sim r^0$ & Yes & No \\ \hline
$\{f_1,f_3,f_6,f_8\}$ & $\sim r^{-6}$ & $\sim r^{-6}$ & No & Yes \\ \hline
  \end{tabular}
\end{table}

The power-law asymptotics are in fact exact solutions in this case.
We see that the asymptotic behaviour at infinity did not depend on $r_0$, which is obvious from
(\ref{eq:fns}) anyway. Whilst the behaviour at the origin does change, it is qualitatively the same and
still none of the solutions are both $L^2$ at the origin and localised
at the origin. The possibility of the Lichnerowicz theorem not holding
existed in this case because the manifold is singular.

\section{Non-radial solutions}

Some non-radial solutions may be obtained by using relationships between eigenfuntions of different
differential operators. This is possible because of the existence of covariantly constant spinors on
special holonomy manifolds \cite{jo}. The idea is developed in \cite{hp} for four dimensional
hyperK\"ahler metrics and in \cite{gpp} for massive eigenfunctions on $G_2$ and Spin(7) metrics.
The massless case we need is slighty more subtle. We will find solutions for the massles Dirac equation
from solutions to the massless scalar equation.
Suppose that $\phi$ satisfies
\be\label{eq:scalar}
\Delta \phi = - \nabla_a\nabla^a \phi = 0 ,
\end{equation}
and that $\eta$ is the covariantly constant spinor, $D\eta=0$.
Then
\bea\label{eq:soluts}
\mbox{if}\quad (\nabla_a \phi) \C^a \eta = 0 & \mbox{then} & \psi = \phi
\eta \quad \mbox{solves}\quad \C^a \nabla_a \psi = 0 , \nonumber \\
\mbox{if}\quad (\nabla_a \phi) \C^a \eta \neq 0 & \mbox{then} & \psi =
(\nabla_a \phi) \C^a \eta  \quad\mbox{solves}\quad \C^a \nabla_a \psi = 0 .
\eea
One can also obtain solutions of the Dirac equation from solutions
of the massless vector equation \cite{gpp},
but this is no easier to solve than the Dirac equation itself and so
is not useful to us here.

The first thing we need to calculate is the covariantly constant spinor. This can be done \cite{gpp, clp}
by imposing the projection conditions
\be
\left( \C_{04} - \C_{23} \right) \eta = \left( \C_{05} - \C_{31} \right) \eta = \left( \C_{06} - \C_{12} \right) \eta = 0 ,
\end{equation}
and then requiring $\eta$ to have constant norm $\eta^{\dagger} \eta = 1$ and to be Majorana as in
(\ref{eq:real}). This first condition is consistent because $D(\eta^{\dagger} \eta) = 0$ and the reality condition
is necessary because the supersymmetry generator must be real for degrees of freedom to match up in the
supergravity theory. All this determines $\eta$ up to a sign, which can be chosen. One obtains
\be
\eta^T = \frac{1}{2} \left( 1, 0, -1, 0, 0, 1, 0, 1 \right) .
\end{equation}
So the covariantly constant spinor has constant components. This will be generically true for cohomogeneity one metrics.
The immediate consequence is that all $r$ dependence of $\psi$ will come from the scalar solution $\phi$.
By substitution it is easy to check that $D\eta = 0$ as required (cf. the constant solutions of \S 2.1 and \S 2.2 above).

The massless Klein-Gordon equation on the manifold can be written
\be
\Delta \phi = -\frac{\delta^{a b}}{\sqrt{g}} \partial_{\mu}
\left( e_{a}^{\mu} e_{b}^{\nu} \sqrt{g} \partial_{\nu} \phi \right) = 0 ,
\end{equation}
where $e_{a}^{\mu}$ are the inverse vielbeins. This expression for the
Hodge-de Rham operator is well suited for exploiting
the symmetries of cohomogeneity-one metrics (\ref{eq:metric}). It has been used to study the Schr\"odinger
equation on the Atiya-Hitchen metric in \cite{gm} and on the Eguchi-Hanson metric in \cite{mig}. For the metric
(\ref{eq:metric}), the equation becomes
\be\label{eq:KG}
\frac{1}{r^6 (1-r_0^3/r^3)}\partial_r (r^6 (1-r_0^3/r^3)^2 \partial_r \phi) + \frac{12}{r^2} J_i J_i \phi + 
\frac{3}{r^2}\frac{4-r_0^3/r^3}{1-r_0^3/r^3} J^{\prime}_i J^{\prime}_i \phi + \frac{12}{r^2} J_i J^{\prime}_i \phi = 0 ,
\end{equation}
where $J_i$ are the vector fields dual to $\Si^i$ and the $J^{\prime}_i$ are dual to $\si^i$. It follows that these
vectors satisfy the algebra $[J_i, J_j ] = \e_{ijk} J_k$ and
$[J^{\prime}_i, J^{\prime}_j] = \e_{ijk} J^{\prime}_k$. This means that as differential
operators, $J_i J_i$, $J^{\prime}_i J^{\prime}_i$ and $J_i J^{\prime}_i$ are simultaneously diagonalisable. Further, their
eigenvalues are well known from elementary angular momentum theory. Acting on the simultaneous
eigenfunctions $\Phi^{mn}_{st}$ we have
\be
J_i J_i \Phi^{mn}_{st} = -m(m+1) \Phi^{mn}_{st} ; \quad J^{\prime}_i J^{\prime}_i \Phi^{mn}_{st} = -n(n+1) \Phi^{mn}_{st} ; \quad
J_i J^{\prime}_i \Phi^{mn}_{st} = - s t \Phi^{mn}_{st} ,
\end{equation}
where $m,n \in \frac{1}{2}\mathbb{Z}^+ \cup \{ 0 \}$ and $\{s,t\} \in \{m,n\} + \mathbb{Z}$
with $-m \leq s \leq m$ and $-n \leq t \leq n$. Note that the half integers are allowed because the group
in question is $SU(2)$ and not $SO(3)$.
We are not interested in the precise functional form of the angular eigenfunctions. Details and references can be found in
\cite{mig}. One now decomposes the function into modes of the form $\phi = \Phi(r) \Phi^{mn}_{st}$.
The equation (\ref{eq:KG}) becomes
\be
\frac{d^2}{dr^2} \Phi(r) + \frac{6 r^2}{r^3-r_0^3} \frac{d}{dr} \Phi(r) - 12 r \left( \frac{m(m+1) + st}{r^3-r_0^3} + n(n+1)
\frac{4r^3-r_0^3}{4(r^3-r_0^3)^2} \right) \Phi(r) = 0 .
\end{equation}
This equation has four regular singular points, at $r^3=r_0^3$ and at
infinity. It does in fact have a solution in terms of hypergeometric
functions, but these are only valid for $r<r_0$ whilst the range of r is
$r \geq r_0$.

This reduction to a single ordinary differential equation
was only possible because of the relatively large symmetry of the metric (\ref{eq:metric}). More general $G_2$ metrics
with less symmetry would have resulted in a Klein-Gordon equation in which the differential operators were
not all simultaneously diagonalisable.

The behaviour at infinity is independent of $r_0$ and we have that as $r\to\infty$
\be\label{eq:limit}
\Phi(r) \sim r^{\frac{-1}{2} \left(5 \pm \sqrt{25+48Q} \right)} ,
\end{equation}
where $Q = m(m+1) + n(n+1) + st \geq 0$. In the radial case, $Q=0$, the two possible behaviours are just the $\mid\psi\mid \sim r^0$
and $\mid\psi\mid \sim r^{-6}$ solutions which we found in the
previous section. For the second solution we use $\mid \psi\mid
\sim \nabla_r \phi \sim \partial_r \phi$ as discussed above.
The case of $Q \neq 0$ will be discussed below. The behaviour at the origin depends on whether
$r_0$ is nonzero.

\subsection{Solutions with $r_0 \neq 0$}

The behaviour as $\bar{r}\to 0$ for the non radial modes depends on whether the integer $n$ is nonzero.
If $n\neq 0$ then as $\bar{r}\to 0$
\be
\Phi(\bar{r}) \sim \bar{r}^{\frac{-1}{2} \left(1 \pm \sqrt{1+4n(n+1)} \right)} .
\end{equation}
The power of $\bar{r}$ is nonzero and therefore
$\mid \psi \mid \sim \bar{r}^{1/2} \partial_{\bar{r}} \phi \sim \bar{r}^{\frac{-1}{2} \left(2 \pm \sqrt{1+4n(n+1)} \right)}$ .
All of these solutions either are not localised or are not square
integrable at the origin. Recall
that the metric goes as $\sqrt{g} \sim \bar{r}$. The effect of having angular dependence on these solutions is to
increase the degree of divergence at the origin of the localised solutions and increase the degree to which
the non-localised solutions increase away from the origin. 

And if $n=0$ then as $\bar{r}\to 0$
\be
\Phi(\bar{r}) \sim k_1 \bar{r}^0 + k_2 \bar{r}^{-1}.
\end{equation}
These correspond to spinors behaving like $\mid \psi \mid \sim \bar{r}^0$ and $\mid \psi \mid \sim \bar{r}^{-3/2}$.
The first does not decay and the second is not $L^2$ at the origin. This set of solutions does not change its radial
behaviour as the angular dependence changes. The behaviour found is
again consistent with the results of the previous section. 

\subsection{Solutions with $r_0 = 0$}

The solution as $r\to 0$ is the same as in the $r\to\infty$ limit of (\ref{eq:limit}). We now
consider the effect of angular dependence. All of the solutions are nonconstant in $r$
and so
\be
\mid \psi \mid \sim \partial_r \phi \sim r^{\frac{-1}{2} \left(7\pm\sqrt{25+48Q} \right)} .
\end{equation}
Again, all of these solutions either are not localised or are not
square integrable at the origin, because $Q \geq 3/4$.
The metric now goes as $\sqrt{g} \sim r^6$. Again, the effect of including angular dependence is to make the
solutions divergent at the origin more divergent and the solutions that grow away from the origin grow faster.

\section{Discussion}

We have found solutions that are constant, solutions decaying
away from the origin and solutions growing away from the origin. None of these
satisified the two properties of being both localised and square integrable at the origin.
The first of these is expected for degrees of freedom emerging as a homology cycle collapses
and the second is necessary for the solution to make sense quantum mechanically.
Allowing angular depedence worsens the situation in all the solutions we found.

The results admit three possible interpretations. Firstly, that the degrees of freedom corresponding to chiral
fermions localised near the singularity correspond to nonradial solutions of the Dirac equation that we have not found.
Secondly, that solving the Dirac equation is not the way to exhibit
these degrees of freedom in any case, or that a modified Dirac equation
including gauge connections is required.
Thirdly, that the $G_2$ manifold asymptotic to a cone over $S^3 \times S^3$ does not give chiral
fermions in the singular limit, unlike the cones over $SU(3) / U(1) \times U(1)$
and $\mathbb{CP}^3$.

The third of these possibilities is the most interesting. It
links nicely with the results from dualities and anomaly cancellation discussed in the introduction.
It also fits with the picture of
extra massless modes coming from collapsed membranes and the fact that the $G_2$ metric asymptotic
to a cone over $S^3\times S^3$ has zero second homology group. Solving
the Dirac equation on other $G_2$ metrics that are known to give
chiral fermions appears to be more challenging, but could lend further support to this
possibility. Another interesting question is to consider the effect
of including gauge field backgrounds in the Dirac equation.

\vspace{1cm}

Many thanks to Gary Gibbons for useful suggestions. Thanks also to Ruben Portugues, James Sparks and
Tibra Ali for discussions. The author is funded by the Sims scholarship.

\end{document}